\documentclass[3p,times,procedia]{elsarticle}
\usepackage{nupha_ecrc}
\usepackage{amsmath,graphicx}

\volume{00}

\firstpage{1}

\journalname{Nuclear Physics A}

\runauth{J.~Blaizot and L.~Yan}

\jid{nupha}

\jnltitlelogo{Nuclear Physics A}

\usepackage{amssymb}
\usepackage[figuresright]{rotating}

\def\E{{\mathcal E}}
\def\P{{\mathcal P}}

\def\L{{\mathcal L}}

\def\tR{\tau_R}

\def\p{{\bf p}}

\def\Eq#1{Eq.~(\ref{#1})}

\def\Fig#1{Fig.~\ref{#1}}

\def\bra{\langle}
\def\ket{\rangle}

\def\be{\begin{equation}}
\def\ee{\end{equation}}
\def\bea{\begin{eqnarray}}
\def\eea{\end{eqnarray}}

\newcommand \beq{\begin{eqnarray}}
\newcommand \eeq{\end{eqnarray}}

\begin{document}

\begin{frontmatter}



%% Instructions from Editor: Please use the following \dochead only in the preprint version (e-print arXiv etc.); 
%% use empty \dochead{} when submitting to Nuclear Physics A!
\dochead{XXVIIth International Conference on Ultrarelativistic Nucleus-Nucleus Collisions\\ (Quark Matter 2018)}
%% e.g. \dochead{17th International Conference on Dynamical Processes in Excited States of Solids}

\title{Fluid dynamics of out of equilibrium
boost invariant plasmas}


\author{Jean-Paul Blaizot}
\address{
	Institut de Physique Th{\'e}orique, Universit\'e Paris Saclay, 
        CEA, CNRS, 
	F-91191 Gif-sur-Yvette, France} 
\author{Li Yan
}
\address{Department of Physics,
McGill University,
3600 rue University
Montr\'eal, QC
Canada H3A 2T8}
\begin{abstract}
We establish a set of equations for moments of the distribution function. 
In the relaxation time approximations, these moments obey a coupled set of equations that can be truncated order-by-order. Solving the equations of moments, we are able to identify an attractor solution that controls a transition from a free streaming fixed point to a hydrodynamic fixed point. In particular, this attractor solution provides a renormalization of the effective value of the shear viscosity to entropy density ratio, $\eta/s$, taking into account off-equilibrium effects.
\end{abstract}


\end{frontmatter}

% \linenumbers

\section{Introduction}
\label{}

%High-energy heavy-ion collisions carried out at RHIC and the LHC have created
%a hot and dense medium -- Quark-Gluon Plasma (QGP). 
The dynamical evolution 
of Quark-Gluon Plasma (QGP) in heavy-ion collisions 
has been found very close to a perfect fluid, with extremely small
viscous corrections~\cite{Kovtun:2004de,Heinz:2013th}. 
The fluidity of QGP can be best understood through the success of
relativistic hydrodynamics, and correspondingly flow signatures that are
well captured by hydro modeling of heavy-ion collisions. 
For instance, the observed elliptic flow $v_2$ fluctuations from the $\sqrt{s_{NN}}=2.76$ TeV 
Pb-Pb collisions is strongly correlated with fluctuations of the geometrical structure 
of initial state, up to a medium response dominated by the system collective
expansion%Quantitatively, the correlation between $v_2$ and initial
%state can be measured in terms of $v_2$ fluctuations, which are compatible with
%the fluctutions of initial ellipticitity
~\cite{Giacalone:2016eyu,Khachatryan:2015waa}. 

%In recent LHC
%experiments, fluctuations of $v_2$ also suggest that 
%QGP fluid has been created in the high-multiplicity p-Pb collisions.
%Fluid behavior
%of QGP, and t
The application of hydrodynamics requires 
a fast thermalization process during the very early stage of heavy-ion collision.
However, it is a theoretical challenge to realize a short time scale (O(1) fm/c) for 
the generated quarks and gluons to evolve towards local thermal
equilibrium or isotropization, especially considering the QGP fluidity in small
colliding systems such as p-Pb~\cite{Khachatryan:2015waa}. 
%One alternative solution is to relax the condition of local thermal
%equilibrium to isotropization, which is recogized when the longitudinal presssure $\P_L$
%is equal to the tranverse pressure $\P_T$,
%\be
%\P_L = \int \frac{d^3 p}{(2\pi)^3 p^0} p_z^2 f(\tau,\vec p_\perp,p_z)\,,\qquad
%\P_T = \frac{1}{2}\int \frac{d^3 p}{(2\pi)^3 p^0} p_T^2 f(\tau,\vec p_\perp,p_z)\,.
%\ee
One alternative solution is to extend hydrodynamics to out-of-equilibrium
systems.
%Although isotropization has been found sufficient for the application of hydrodynamics,
%it has been recently realized that hydrodynamics can be applied to 
%out-of-equilibrium systems. 
In this work, we propose a set of $\mathcal{L}$-moments, based on 
which a framework of fluid dynamics can be established in out-of-equilibrium and 
boost-invariant systems.

\section{$\mathcal{L}$-moment and $\mathcal{L}$-moment equations}

\begin{figure}
\begin{center}
\includegraphics[width=0.45\textwidth] {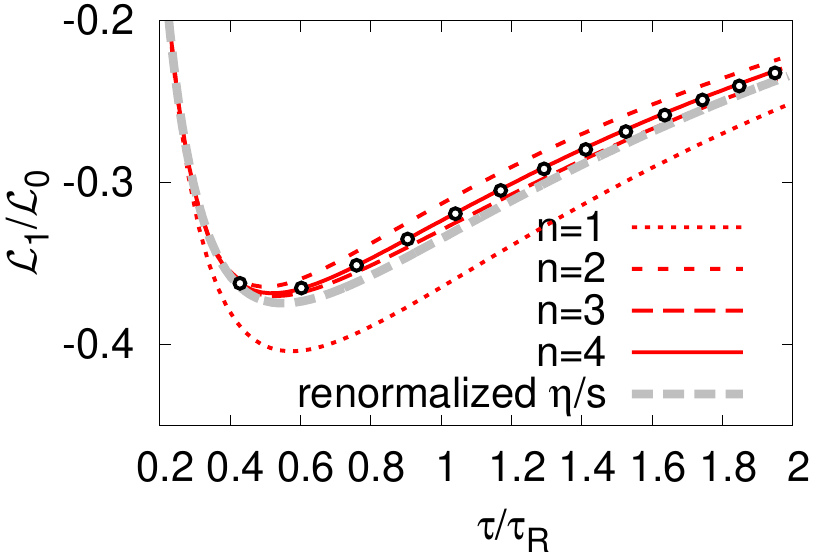}
\caption{(Color online) Comparison of  the $\mathcal{L}$-moment equations obtained from various truncation of Eqs.~(\ref{eq:Leom}) (lines), with 
those of the numerical solution of the kinetic equation (\ref{eq:trans}) (symbols).  
\label{fig:compL}
}
\end{center}
\end{figure}

In the very early stage of high-energy heavy-ion collisions, the evolution
of QGP system
is dominated by the longitudinal expansion along the beam-axis. Accordingly, the
evolution of QGP can be well approximated by Bjorken boost invariance and
one is allowed to write the phase space distribution function 
as $f(\p,\tau)$.
%The phase space distribution defines
%energy density $\epsilon$, pressures $\P$, etc. It is worth
%mentiong that these quantities are related to hydrodynamics, through the definition of energy-momentum tensor,
%\be
%T^{\mu\nu}=\int \frac{d^3 p}{(2\pi)^3 p^0} p^\mu p^\nu f(\tau,\vec p_\perp,p_z)\,.
%\ee 
%One also has
%the longitudinal presssure $\P_L$ and the tranverse pressure $\P_T$,
%\be
%\P_L = \int \frac{d^3 p}{(2\pi)^3 p^0} p_z^2 f(\tau,\vec p_\perp,p_z)\,,\qquad
%\P_T = \frac{1}{2}\int \frac{d^3 p}{(2\pi)^3 p^0} p_T^2 f(\tau,\vec p_\perp,p_z)\,.
%\ee
%which are used to characterize the isotropization condition: $\P_L=\P_T$. 
%
Given the phase space distribution, we introduce the $\mathcal{L}$-moment~\cite{Blaizot:2017lht,Blaizot:2017ucy},
\be
\label{eq:Lndef}
\mathcal{L}_n = \int \frac{d^3 p}{(2\pi)^3 p^0} p^2 P_{2n}(p_z/p_\perp)
f(\p,\tau)\,,
\ee
where $P_{2n}(x)$ is Legendre polynomial of order $2n$. $\L$-moments are of the same
dimension as the energy-momentum tensor $T^{\mu\nu}$, but contains more detailed information of
the anisotropic structure of $f$. One may check that 
the two lowest order moments, $\mathcal{L}_0$ and $\mathcal{L}_1$,
coincide with energy density and pressure anisotropy, respectively.
%$\L_0=\epsilon$ and $\L_1=\P_L-\P_T$.
A vanishing $\mathcal{L}_1$ corresponds to isotropization.
To derive the equations of motion for $\mathcal{L}_n$, for simplicity, 
we consider a transport equation with relaxation time approximation for the boost-invariant
QGP,
\be
\label{eq:trans}
\left[\partial_\tau - \frac{p_z}{\tau}\partial_{p_z}\right] f(\p,\tau) = -\frac{f(\p,\tau)-f_{\rm eq}(p/T)}{\tau_R},
\qquad\tR=\tR(T)%\sim \frac{\eta}{s}
\ee
where the relaxation time $\tR$ is a function of local temperature. Especially, throughout
this paper, we consider a 
conformal system, $\tR\propto \eta/sT$. As will become clear later, the dimensionless quantity
$\tau/\tR$ characterizes inverse of Knudsen number of the expanding system.

\Eq{eq:trans} leads a set of coupled equations for $\mathcal{L}_n$,
\be
\label{eq:Leom}
\frac{\partial \mathcal{L}_n}{\partial \tau} = -\frac{1}{\tau}\left[a_n\mathcal{L}_n
+b_n\mathcal{L}_{n-1}+c_n\mathcal{L}_{n+1}\right]-\frac{\mathcal{L}_n}{\tR}(1-\delta_{n0}),\quad n=0,1,...
\ee
with $a_n$, $b_n$ and $c_n$ constant coefficients from the recursion relations of Legendre
polynomials. One has to truncate \Eq{eq:Leom} for practical analyses. A straightforward way to
truncate \Eq{eq:Leom} at order $n$ is to ignore all $\mathcal{L}$-moments higher than the n-th one.
Especially, we notice that truncation at $n=0$ gives ideal hydro equation of motion.
A test of the truncation is shown in \Fig{fig:compL}. With respect to 
the evolution of $\mathcal{L}_1/\mathcal{L}_0$, from the case of truncation at $n=1$ to the case of
truncation at $n=4$, one indeed observes a trend of convergence towards the exact solution (open
symbols).

\section{Fixed points of $\mathcal{L}$-moment equations and hydro attractors}

%\begin{figure}
%\begin{center}
%\includegraphics[width=0.40\textwidth] {./figs/att_eta_1}
%\caption{ Renormalization constant $Z_{\eta/s}$ as a function of $\tau/\tau_R$. The leading order corresponds to Eq.~(\ref{L2L1}), the next-to-leading order include the correction due to $g_3(\tau)$. 
%\label{fig:att}
%}
%\end{center}
%\end{figure}
%

There exist fixed-point solutions of \Eq{eq:Leom} in two extreme cases. 
These fixed points are analyzed in terms of the following dimensionless quantity,
\be
g_n = \tau \partial_\tau \ln \mathcal{L}_n\,,
\ee
which characterizes the decay rate of $\mathcal{L}$-moment regarding the system expansion.
The first extreme with
$\tau/\tR\rightarrow0$ corresponds to the free-streaming evolution. In the free-streaming case,
which amounts to very early time or sufficiently long relaxation time (or small coupling) 
in the expanding QGP system,
one finds two fixed points: $g_n\approx -2$ and $g_n\approx -1$. Several comments are in order.
First, it is not difficult to verify that only the one around $-1$ is a stable fixed point, which
reflects the one-dimensional expansion of the system.
Secondly, fixed points of all orders of $\mathcal{L}$-moments degenerate: $g_0=g_1=\ldots$.
Lastly, the fixed points are found exact at $-2$ and $-1$ without truncation of the coupled
$\mathcal{L}$-moment equations. However, truncation at any finite order will vary the expected value of fixed points. 
For intance, instead of -1, the stable fixed points are $g_0=g_1=-0.92937$ when truncating at $n=1$.

The opposite extreme $\tau/\tR\rightarrow \infty$ corresponds to the hydro limit, in which the
system evolution reduces to hydrodynamics. To evaluate $g_n$ in this limit, we notice that the coupled equations
of the $\mathcal{L}$-moments can be identified with hydro equation of motion. For truncation $n=0$, it trivially
leads to the ideal hydro equation of motion. For the truncation at $n=1$, the coupled equations for $\mathcal{L}_0$
and $\mathcal{L}_1$ lead to the second order viscous hydro (Israel-Stewart), 
\begin{align}
\label{eq:hydro}
&\partial_\tau \mathcal{L}_0=-\frac{1}{\tau}(a_0\mathcal{L}_0 +c_1\mathcal{L}_1)\quad\rightarrow\quad
\partial_\tau \E+\frac{\E+\P_L}{\tau}=0\cr
&\partial_\tau \mathcal{L}_1=-\frac{1}{\tau}(a_1\mathcal{L}_1 + b_1\mathcal{L}_0)-\frac{1}{\tR}\mathcal{L}_1\quad\rightarrow\quad
%\partial_\tau \Pi = -\frac{b_1 c_0 \varepsilon}{\tau}-\frac{a_1\Pi}{\tau}-\frac{\Pi}{\tR}
\Pi=-\eta\sigma - a_1 \frac{\tR}{\tau}\Pi-\tR \partial_\tau\Pi\,.
\end{align}
as long as one identifies $c_0\mathcal{L}_1=\Pi=\Pi^\xi_{\;\xi}$ as the $\xi$-$\xi$ component of the
shear stress tensor. 
In \Eq{eq:hydro}, $\sigma=\sigma^\xi_{\;\xi}$ is the $\xi$-$\xi$ component of the tensor
$\sigma^{\mu\nu}=2\bra\nabla^\mu u^\nu\ket$.
Similar strategy can generalized to truncation at arbitrary
orders, and one recovers hydro equation of motion of higher orders. In these derivations, we notice
that the leading order term in the gradient expansion of $\mathcal{L}_n$ is $\mathcal{L}_n\sim 1/\tau^n$, from which
we find in the hydro limit, $g_n =-\frac{4+2n}{3}$. Note that these are stable fixed points depending on order $n$.

If initially starting from the stable free-streaming fixed point ($g_n\approx-1$), the system will
evolve and finally approach to the hydro fixed points. 
These are special solutions of the system evolution, since any variation of these solutions will
damp quickly due to the properties of fixed points, hence they are hydro attractors.
Hydro attractor has been studied recently in various aspects, with respect to hydrodynamics~\cite{Heller:2015dha},
kinetic theory~\cite{Romatschke:2017vte}, and systems beyond Bjorken symmetry~\cite{Denicol:2018pak}. 
Numerically we have confirmed that hydro attractors identified between 
fixed points are consistent with what was found by
other methods~\cite{Romatschke:2017vte}. It should be emphasized that, in addition to the lowest order moment, 
$g_0 \sim (\P_L-\P_T)/\epsilon$, there are infinite number of attractors in the evolving system.

\section{Hydro attractors and renormalization of $\eta/s$}

\begin{figure}
\begin{center}
\includegraphics[width=0.45\textwidth] {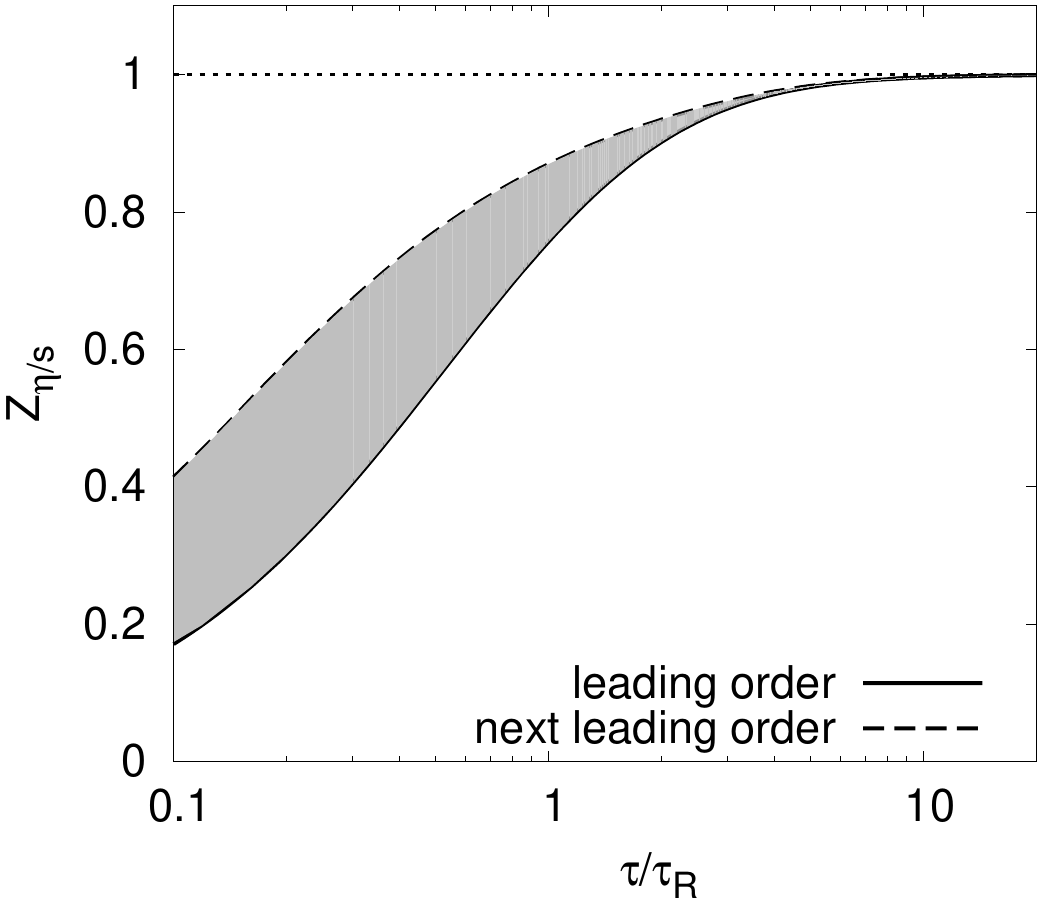}
\caption{ Renormalization constant $Z_{\eta/s}$ as a function of $\tau/\tau_R$. 
%The leading order corresponds to Eq.~(\ref{L2L1}), the next-to-leading order include the correction due to $g_3(\tau)$. 
\label{fig:etasR}
}
\end{center}
\end{figure}

Considering the fact that attractor solutions extend the description of system evolution to
out of equilibrium, we apply our hydro attractors to the formulation of out-of-equilibrium 
hydrodynamics. We understand that truncation
at $n=2$ gives third order viscous hydrodynamics. To retain
effects related to third or higher order viscous corrections, we rewrite the coupled equations as
\begin{align}
\label{eq:ren}
\partial_\tau \mathcal{L}_0 =& -\frac{1}{\tau}(a_0\mathcal{L}_0+c_0 \mathcal{L}_1)\,,\cr
\partial_\tau \mathcal{L}_1 =& -\frac{1}{\tau}(a_1\mathcal{L}_1+b_0 \mathcal{L}_0)
-\left[1+\frac{c_1\tR}{\tau}\frac{\mathcal{L}_2}{\mathcal{L}_1}\right]\frac{\mathcal{L}_1}{\tR}
\end{align}
where the factor in the brackets (we define as $Z_{\eta/s}^{-1}$) 
is related to $g_2(\tau/\tR) = -a_2-b_2\frac{\mathcal{L}_2}{\mathcal{L}_1} - \frac{\tau}{\tR}$. 
If one takes the attractor solution for $g_2$, higher order viscous effects are absorbed into $g_2$
through gradient summation.
Accordingly, in \Eq{eq:ren},
the factor $Z_{\eta/s}^{-1}$ effectively renormalizes the relaxation time $\tR$, or $\eta/s$.
\Fig{fig:etasR} presents the factor $Z_{\eta/s}^{-1}$ based on numerical solution of the 
hydro attractor, with respect to truncation at $n=3$ (leading order) and $n=4$ (next leading order).
One notices that for system close to equilibrium ($\tau/\tR\rightarrow \infty$) the off-equilibrium effects
are small, as expected. On the other hand, when the system is far away from equilibrium ($\tau/\tR\rightarrow0$),
the effective value of $\eta/s$ is largely reduced. The effect of $\eta/s$ renomralization can be further
examined in practical simulations. A preliminary result is shown in \Fig{fig:compL} as the grey dashed line, 
which is solved numerically from the second order hydrodynamic equation but with a renormalized $\eta/s$.
Comparing to the solution from second order hydro (truncation at $n=1$ in \Fig{fig:compL}), improvement
is remarkable.

\section{Summary}

We have proposed a set of $\mathcal{L}$-moments, whose equations of motion can be applied to describe out-of-equilibrium 
system evolution. These equations coincide with hydro equations of motion in the hydro regime. Hydro attractors
can be found with respect to the $\mathcal{L}$-moments, and an effective renormalization of $\eta/s$ can be derived, which
contains effects of out-of-equilibrium dynamics.

\section*{Acknowledgements}
LY is supported in part by the Natural Sciences and
Engineering Research Council of Canada.



%% References


\bibliographystyle{elsarticle-num}
\bibliography{refs-qm18.bib}






% \end{thebibliography}

\end{document}